\documentclass[cits]{PoS}
\usepackage{graphics}
\newcommand{\mdot}{\dot{M}}
\def\gta{ \lower .75ex \hbox{$\sim$} \llap{\raise .27ex \hbox{$>$}} }
\def\lta{ \lower .75ex\hbox{$\sim$} \llap{\raise .27ex \hbox{$<$}} }

\title{A unified  paradigm for the spectral and temporal evolution of Black Hole X-ray Binaries}

\ShortTitle{A Unified paradigm for BHXrB}

\author{\speaker{P.-O. Petrucci}, J. Ferreira, C. Cabanac, G. Henri and G. Pelletier\\
        LAOG, Grenoble, France\\
        E-mail: \email{pierre-olivier.petrucci@obs.ujf-grenoble.fr}}

\abstract{We present a unified picture to explain the spectral states of BH
binaries based on the two-flow model developed in our team since several years and already applied with success to AGNs. In our view, the central regions have a multi-flow
configuration consisting in (1) an outer standard accretion disc (SAD)
down to a transition radius $r_{tr}$, (2) an inner magnetized jet emitting disc (JED)  below $r_{tr}$
driving (3) a self-collimated non relativistic MHD jet surrounding, when
conditions for pair creation are met, (4) a ultra relativistic pair
beam. Large values of $r_{tr}$ correspond to Hard states while small values
correspond to Soft states. In between these extremes, in the high
intermediate state, $r_{tr}$ can reach values that switch on a pair cascade
process giving birth to ultra-relativistic pair blobs that explain the
superluminal events.  In this model the accretion rate but also the disc magnetization $\mu$ play important roles, the latter being necessarily in the range 0.1--1 for the MHD jet to exist. Then, with simple assumptions on $\mu$,  we propose an explanation for  the hysteresis  behavior observed in microquasars during their outburst .  We also discuss the nature, in our framework, of the X-ray corona. While it is commonly believed to be at the base of the jet it cannot be in the jet itself, its optical depth being to small to be consistent with the X-ray spectra observed in the Hard state. We show that it is expected to form naturally in the JED itself.  Finally we investigate the nature of the X-ray variability in the hard state  by assuming an oscillating hot plasma whose
temperature and density vary locally through the propagation of a magneto-sonic
wave. The variable comptonized spectra are computed through Monte-Carlo
simulation. Preliminary results are discussed.}

\FullConference{VI Microquasar Workshop: Microquasars and Beyond\\
		September 18-22 2006\\
		Societ\`a del Casino, Como, Italy}

\begin{document}
\section{The two-flow model}
The two-flow model consists in combining a relativistic flow along the axis of a non-relativistic or mildly relativistic 
MHD flow coming from the accretion disk (cf. Fig. \ref{fig:smae2flow}). When this model has been proposed in the eighties (mostly with a description 
of the interaction of the cold relativistic beam with an ambient medium, e.g. \cite{pel88b}), the arguments were rather poor: 
the phenomenology of hot spots do not require a relativistic flow, whereas relativistic sporadic ejections are observed 
only at short distance from the source; the large scale jet does not seem to lose significant energy during its travel, 
and non-relativistic shock acceleration satisfactorily explains the hot spot synchrotron spectra. So the large scale 
non relativistic-jets and the short scale relativistic-jets could be generated differently. First astrophysical applications of the model have been
published in 1989 \cite{sol89,pel89,pel92}. But more arguments have been proposed later \cite{hen91,hen93} in relation with X-ray 
and gamma ray emissions. The interest of the two-flow model relies on the following main arguments:
\begin{enumerate} 
\vspace*{-0.2cm}\item The relativistic jet alone would radiatively cool rapidly and be relaxed by Compton drag \cite{phin87}. The heating of the relativistic jet particles through MHD wave interaction with the non-relativistic flow suppresses or at least strongly weakens this effect.
\vspace*{-0.4cm}\item The relativistic jet does not self-collimate \cite{bog01b,pel04}. The non-relativistic MHD flow, which self-collimates, can confine it and 
manages an ecological niche for it, because of the centrifugal barrier inherently associated with the MHD acceleration 
of the non relativistic-jet.  
\vspace*{-0.4cm}\item The extraction of rotation energy from a spinning Black Hole does not generate a power strong enough to 
account for the large power of FR2 jets. Jets powered by accretion are more powerful, but are mildly relativistic \cite{pel04}. 
\vspace*{-0.4cm}\item The relativistic-flow is likely in the form of flaring clouds, rich in electron-positron pairs, channelled by the non relativistic-jet, and involving 
a smaller power. It probably does not travel beyond kpc distance from its source.
\end{enumerate}

\vspace*{-0.2cm}A new version of the two-flow model has been developed since 1991, where the relativistic-jet is a hot pair plasma, energized 
by turbulence from the non relativistic-jet \cite{hen91, hen93} and experience a Compton Rocket effect. Satisfactory fits of the spectra of 
3C273 and 3C279 has been obtained with a natural break at a few MeV \cite{mar95}; the stratification, which appears hardly avoidable  (e.g. \cite{hen06}) plays a major role in 
the analysis, which deeply differs from homogeneous or slab models . Later, a study of the micro-instabilities of such a 
relativistic-jet have been investigated \cite{mar97}; these instabilities differ from the one investigated in the previous version, by the 
fact that they are triggered by the streaming of the cold ambient medium in the hot relativistic-jet, instead of considering the 
relativistic-jet as a cold beam exciting waves by pervading the ambient medium. This is an interesting circumstance of plasma 
turbulence excitation, because the streaming instability is driven by the radiation force and quasi-linear theory is 
not sufficient to describe the stationary state; non-linear mode couplings are necessarily taken into account, and, in 
particular, allow to estimate the heating of the pair plasma. 

\section{Application of the two-flow model to microquasars}

The application of the two-flow model to microquasars is relatively natural. First, there is a large consensus on the fact that the physical processes occurring in AGNs and microquasars are similar, the major difference being the global size of the system which depends linearly on the central black hole mass (from tens of solar masses in microquasars to millions or even billion of solar masses in AGNs). Then, it appears reasonable to apply a model that has been validated in the AGNs framework (cf. previous section) to the microquasars one. On the other hand, the importance of ejection processes in microquasars as well as the crucial links between accretion and ejections during their spectral state evolutions  (deduced from accurate multi-wavelength studies, e.g. \cite{cor04,fen04c}) imply that accretion and ejection processes have to be addressed all together in a global and consistent way. This is an important characteristic of our model where accretion and ejection are indeed treated in a self-consistent way (cf. below). The first  self-similar solution of the complete set of equations of an accretion-ejection structure can be found in \cite{fer95,fer97}. Analytical computations and heavy numerical simulations have been developed since that time in different astrophysical contexts \cite{cas00a,cas00b,fer04,pes04,cas04,fer06b} and their validity have been confirmed by different authors (e.g.  \cite{kon04,zan06}). Finally, there are growing evidences of the presence of different jet velocities component in XrB, one moving at non or mildly relativistic speed while the other propagates in the ultra-relativistic regime \cite{fom01a,fom01b,fen04, migl05a}. While these results do not request univocally the presence of two different jet components, they are of course naturally explained by our model.

\subsection{General picture}
\label{mainpict}
\begin{figure*}
\centering
\includegraphics[width=0.6\textwidth]{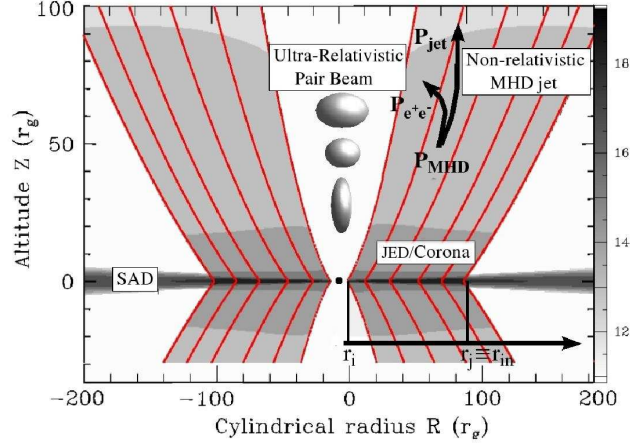}
\caption{Application of the two-flow model to microquasars. This is a sketch of our model showing its different dynamical components. A Standard Accretion Disc (SAD) is established down to a radius
$r_{tr}$ which marks the transition towards a low radiative Jet Emitting
Disc (JED), settled down to the last stable orbit. The JED is driving a
mildly relativistic, self-collimated electron-proton jet which, when
suitable conditions are met, is confining an inner ultra-relativistic
electron-positron beam. The MHD power $P_{MHD}$ flowing from the JED acts
as a reservoir for (1) heating the inner pair beam
($P_{e^+e^-}$) and (2) driving the compact jet ($P_{jet}$). We also assume the presence of a thermal hot corona at the base of the jet, and which is more likely part of the JED itself (cf. Sect. 4). Field lines
are drawn in black solid lines and the number density is shown in
grayscale ($\log_{10} n/\mbox{m}^{-3}$).}
\label{fig:smae2flow}
\end{figure*}

We assume that the central regions of BH XrB are composed of four
distinct flows: two discs, one outer "standard" accretion disc (hereafter
SAD) and one inner jet emitting disc (hereafter JED), and two jets, a
non-relativistic, self-confined electron-proton MHD jet and, when
adequate conditions for pair creation are met, a ultra-relativistic
electron-positron beam. We also assume the presence of a hot thermal corona at the base of the jet that is more likely more likely part of the JED itself (cf. Sect. 4).  A sketch of our model is shown in
Fig.~\ref{fig:smae2flow} and the four dynamical components are
discussed separately in more details in \cite{fer06a}. This
model provides a promising framework to explain the canonical spectral
states of BH XrBs mainly by varying the transition radius $r_{tr}$ between
the SAD and the JED. This statement is not new and has already been
proposed in the past by different authors
(e.g. \cite{esi97,bel97,liv03,kin04}) but our model distinguishes
itself from the others by the consistency of its disc--jet structure and
by the introduction of a new physical component, the ultra-relativistic
electron-positron beam, that appears during strong outbursts.

The electron-proton MHD jet is collimated by large
scale vertical field $B_z$ anchored somewhere in the accretion disc (the JED)
and we assume that this large scale magnetic field has the same polarity everywhere. The
presence of a large scale vertical field threading the disc is however
not sufficient to drive super-Alfv\'enic jets. This field must be close
to equipartition as shown by \cite{fer95} and \cite{fer97}. The reason is
twofold. On one hand, the magnetic field is vertically pinching the
accretion disc so that a (quasi) vertical equilibrium is obtained only
thanks to the gas and radiation pressure support. As a consequence, the
field cannot be too strong. But on the other hand, the field must be
strong enough to accelerate efficiently the plasma right at the disc
surface (so that the slow-magnetosonic point is crossed smoothly). These
two constraints can only be met with fields close to equipartition.

An important local parameter is therefore the disc magnetization $\mu =
B_z^2/(\mu_o P_{tot})$ where $P_{tot}$ includes the plasma and radiation
pressures. In our picture, a SAD is established down to a radius $r_{tr}$
where $\mu$ becomes of order unity. Inside this radius, a JED with $\mu
\sim 1$ is settled.  At any given time, the exact value of $r_{tr}$ depends
on highly non-linear processes such as the interplay between the amount
of new large scale magnetic field carried in by the accreting plasma
(e.g.. coming from the secondary) and turbulent magnetic diffusivity
redistributing the magnetic flux already present. These processes are far
from being understood. For  sake of simplicity, we will treat in this
section $r_{tr}$ as a free parameter that may vary with time  independently of the accretion rate (but see
Section~3 for a more realistic picture). In that respect, our view is
very different from that of \cite{esi97,mah97} who considered only the
dependency of the accretion rate  $\mdot$ to explain the different spectral states of BH
XrBs.

\subsection{Time evolution of BH XrBs}
\label{evolution}
The evolution with time of a BH XrB has been reported in
Fig.~\ref{fig:hid}. This is a synthetic
Hardness--Intensity diagram (hereafter HID) as it is generally observed
in XrBs {in outbursts} (e.g. \cite{bel05,fen04c}) .  { During such outbursts, the objects follow the A-B-C-D sequence shown on the figure 
before turning back to A at the end of the outburst}.  We have
over-plotted on Fig. \ref{fig:hid} the different sketches of our
model (this figure is clearly inspired by Fig. 7 of \cite{fen04c}) and we have plotted on Fig. \ref{diffstate}  the corresponding SEDs expected in each spectral state. We detailed below
the interpretation of the HID in our framework. \\

\begin{itemize}
\item {\bf Ascending the Right Branch:} Let us start at a Low/Hard State (LHS) located at the bottom of the HID right
branch (point A in Fig. \ref{fig:hid}). Such state would
correspond to a JED extending up to typically $r_{tr} \sim 10^2 $ (in units of gravitational radii $r_G$).  This
considerably lowers the emission from the inner radii of the SAD
producing a weak UV/soft X-ray excess. The hard (1-20 keV) power-law component
of photon index $\Gamma \sim 1.7$ is attributed to the warm thermal
plasma at the base of the jet. The non relativistic MHD jet then produces
the persistent IR and optically thick radio synchrotron emission (cf. the corresponding SED in Fig. \ref{diffstate}).
\item {\bf The Top Horizontal Branch: } 
\begin{figure*}
\centering
\begin{tabular}{cc}
\includegraphics[width=0.6\textwidth]{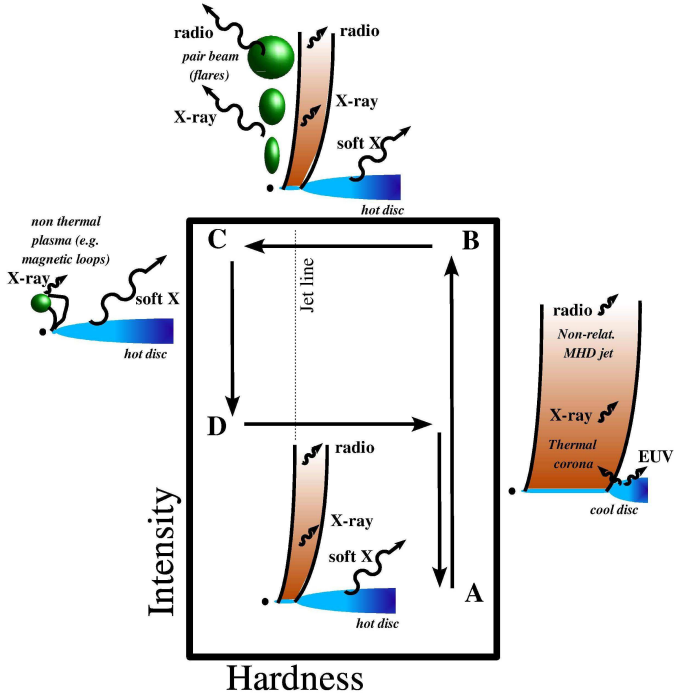}&
\begin{minipage}{0.33\textwidth}
\vspace*{-13cm}
\caption{Schematic Hardness--Intensity diagram as it is generally
  observed in XrBs in outbursts (this figure is clearly inspired by
  Fig. 7 of \cite{fen04c}). During such outbursts, the objects
  follow the A-B-C-D sequence before turning back to A at the end of the
  outburst. The jet line has been defined by \cite{fen04c} and separates jet-dominated states to disc dominated ones. See Sect. 2.2 for more details. \label{fig:hid}}
\end{minipage}\\
\end{tabular}
\end{figure*}
\begin{itemize}
\item {\underline{Before the jet line :}}  {Arriving in B} we assume that $r_{tr}$ starts decreasing rapidly. Then, the JED undergoes an outside-in transition to a SAD. The BH XrBs enter
the high intermediate state. The flux of the outer standard disc then increases while the JED is
decreasing. Under such circumstances, the MHD Poynting flux released by
the JED is still important (through the large $\mdot$ that characterizes
this part of the HID) but the accretion-ejection structure (i.e. JED and MHD jet) itself fills a smaller volume, a
direct consequence being a weaker emission of the thermal "corona" and
the non-thermal MHD jet emission with respect to what it is while in the LHS.
\begin{figure*}
\centering
\begin{tabular}{cc}
\includegraphics[width=0.6\textwidth]{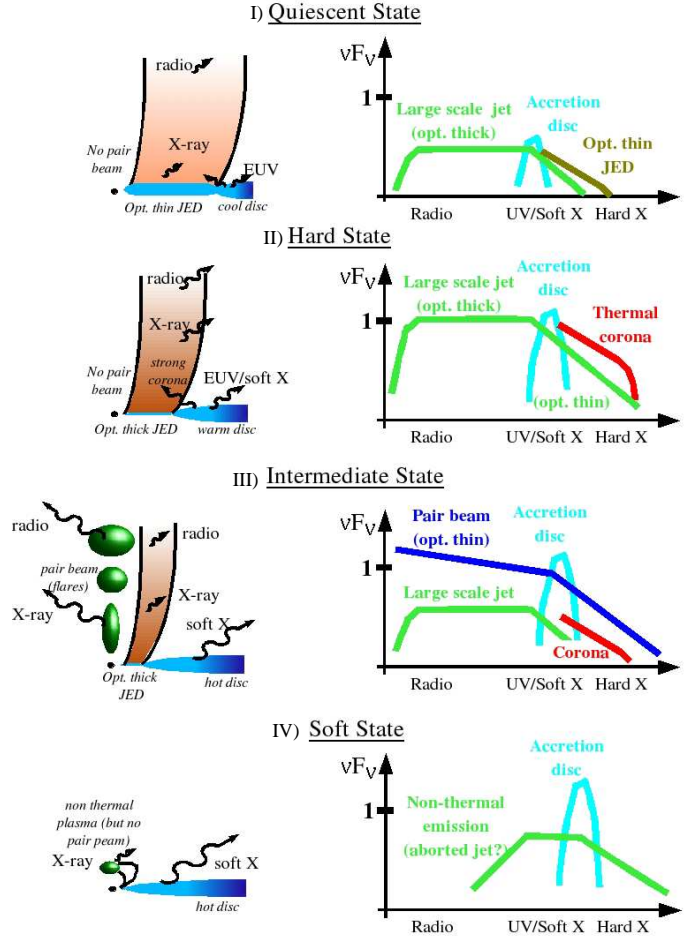}&
\begin{minipage}{0.33\textwidth}
\vspace*{-13cm}
\caption{The canonical spectral states of X-ray binaries
 . {\bf (I)} Quiescent state
  obtained with a low $\mdot$ and a large $r_{tr}$: the Jet Emitting Disc
  (JED) occupies a large zone in the accretion disc. {\bf (II)} Hard state
  with much larger $\mdot$ and smaller $r_{tr}$: the pair creation threshold
  is not reached yet. {\bf (III)} Luminous Intermediate state between the
  Hard and the Soft states: the high disc luminosity combined with the
  presence of a MHD jet allows efficient pair creation and relativistic acceleration along the
  axis, giving birth to flares and superluminal ejection events.{\bf (IV)}
  Soft state when there is no zone anymore within the disc where an
  equipartition field is present: no JED, hence neither MHD jet nor pair
  beam.\label{diffstate}}
\end{minipage}\\
\end{tabular}
\end{figure*}
\item {\underline{At the jet line:}} The jet line has been defined by \cite{fen04c} and separates jet-dominated states to disc dominated ones.  This correspond to the critical phase reaches by the system during its evolution along the top horizontal branch where the conditions for a strong pair production are fulfilled (cf. \cite{fer06a} for more details). In this case, we expect an explosive behavior of the pairs, with the
sudden ejection of blobs. The emission of these blobs, first in X and
$\gamma$ ray and then, after a rapid expansion, in IR and radio, will
probably dominate the broad band spectrum, producing the hard X-ray tail
and the optically thin radio emission present in this state (cf. the corresponding SED in Fig. \ref{diffstate}). The
production of a series of blobs can even result in an apparently
continuous spectrum, from radio to X/$\gamma$ rays.
We note that, at the jet line,  the rapid increase of the pair beam pressure in the inner
region of the MHD jet, during e.g. a strong outburst, may dramatically
perturb the MHD jet production and even makes it disappear in agreement with observations \cite{fen04c}.
It is also worth noting that  we do not expect all BH
XrBs to reach a ``pair production'' phase along the top branch. Indeed, triggering the pair cascade process requires a large enough  gamma-gamma opacity and hence a high accretion rate that could not be reached by all objects.

\item {\underline{After the jet line:}} We assume that $r_{tr}$ is still decreasing. We therefore expect the total
disappearance of the JED and its MHD jet when $r_{tr}$  approaches the disc inner radius $r_{in}$,
thereby also causing the end of the pair beam { (if present)}. The inner
regions of the BHXrB are a SAD with probably a strong magnetic activity. Indeed, the situation might be slightly
more complex than a mere SAD because of the presence of a concentrated
magnetic flux. No steady MHD ejection can be produced from the SAD but
unsteady events could be still triggered. This is maybe the reason why
this region in the HID seems to harbor complex variability phenomena
\cite{bel05,nesp03}.

\end{itemize}

\item {\bf Descending the Left Branch:}  When XrBs reach the left vertical branch (point C in
Fig. \ref{fig:hid}), $r_{tr}$ is smaller than the inner disc radius $r_{in}$ i.e. the
JED and the MHD jet have completely disappeared.  The whole disc adopts
therefore a radial structure akin to the standard disc model and we enter
into the so-called soft state (also called thermal dominant state by \cite{mcc03})
where the spectra are dominated by strong disc emission (cf. the corresponding SED in Fig. \ref{diffstate}). The descent from
C to D correspond to a decrease in intensity i.e. by a decrease of the
accretion rate. This is the beginning of the fading phase of the
outburst, $r_{tr}$ keeping smaller than $r_{in}$. We note also that we still expect the presence of magnetic fields that
may be the cause of particle acceleration responsible for the weak
hard-energy tail generally observed in this state (\cite{mcc03,zdz04} and references therein).

\item {\bf The Low Horizontal Branch:} In D we assume that $r_{tr}$ begins to increase again. Thus, according to this conjecture,
there is an inside-out build up of a JED. Self-collimated electron-proton
jets could be produced right away. This means an increase of $r_{tr}$, the
reappearance of the non-thermal MHD jet and the thermal corona and a decrease of the SAD emission. But, contrary to the Top
Horizontal Branch, the accretion rate is now too low to allow the
production of a pair beam. { Consequently we do not expect superluminal
motions during this phase.}

\end{itemize}

\section{What can drive $r_{tr}$ ?}
\label{muvarsect}
From the previous sections, it is clear that the spectral shape of a BH XrB
critically depends on the size of the JED relative to the SAD, namely
$r_{tr}$. As stated before, $r_{tr}$ is the radius where the disc magnetization
$\mu = B_z^2/(\mu_o P_{tot})$ becomes of order unity.  Thus, $r_{tr}$
depends on two quantities $P_{tot}(r,t)$ and $B_z(r,t)$. The total
pressure is directly proportional to $\mdot$ since $P_{tot}= \rho
\Omega_k^2 h^2 \propto \mdot M^{-1}r^{-5/2}$, $\rho, \Omega_k, h$ and $M$ being respectively the disc mass density, the keplerian angular velocity, the disc scale height and the black hole mass. As a consequence, any
variation in the outer SAD accretion rate implies also a change in
the amplitude of the total pressure. But we have to assume something
about the time evolution of the large scale magnetic field threading the
disc. The processes governing the amplitude and time scales of these
adjustments of $r_{tr}$ to a change in $\mdot$ are far too complex to be
addressed here. They depend on the nature of the magnetic diffusivity
within the disc but also on the radial distribution of the vertical
magnetic field. As already noted, up to now we have simply assumed  that $r_{tr}$ and
$\mdot$ are two independent parameters. However we can try  to relate these two parameters with each other using simple but  reasonable assumptions. 

\begin{figure}
\centering
\includegraphics[width=0.65\textwidth]{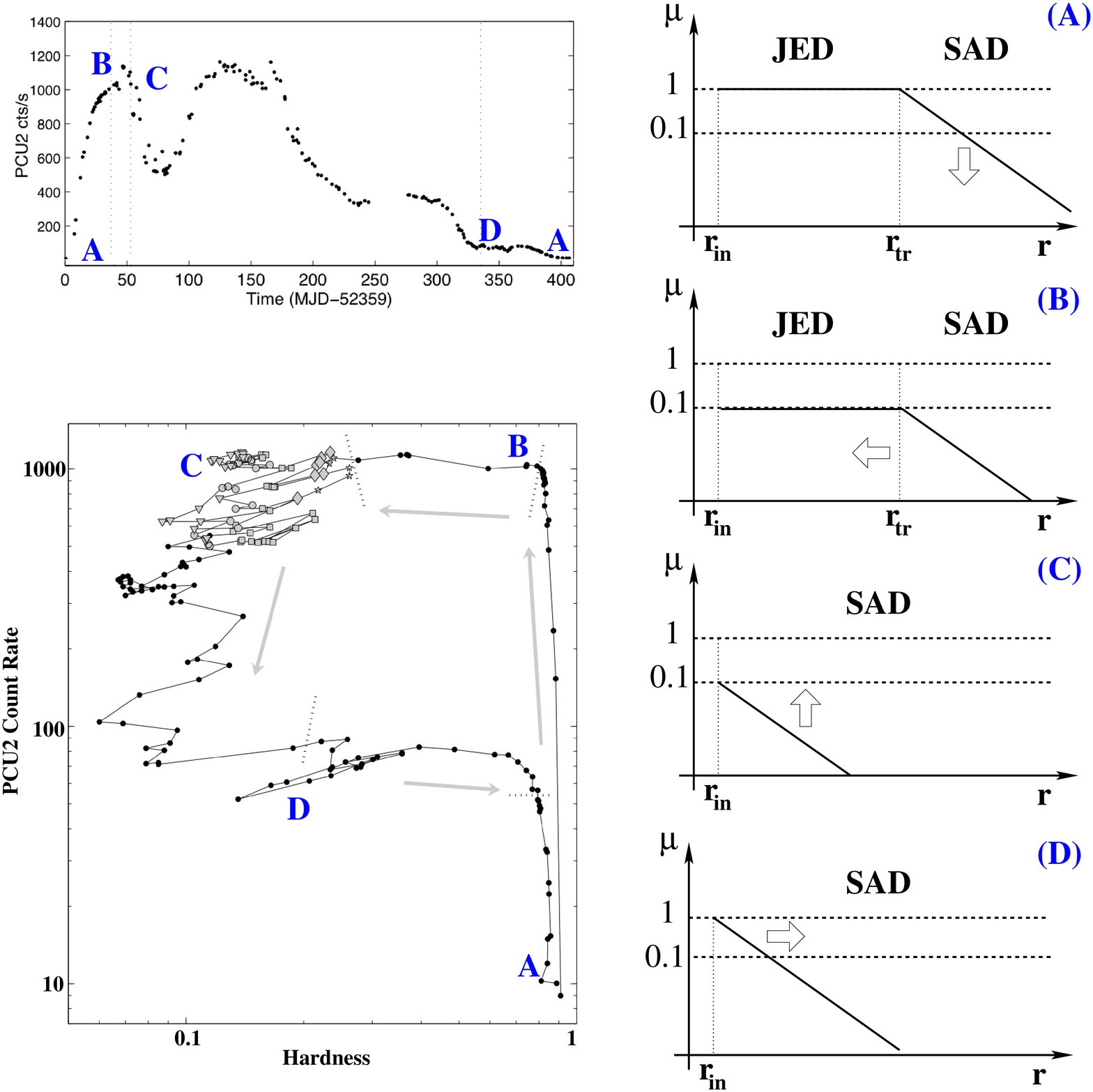}
\caption{The expected evolution of the radial distribution of the disc
  magnetization $\mu$ during an outburst where an hysteresis in the
  Hardness-Intensity diagram is observed. The light curve and HID are
  those followed by GX 339-4 during its 2002-2003 outburst
  \cite{bel05}. See Sect. 3 for details. }
\label{muprofil}
\end{figure}

If the disc magnetization $\mu$ has to be close to unity for an accretion-ejection structure to exist  there is obviously some freedom on the  acceptable value of this parameter. More precisely, analytical and numerical computations show that $\mu$ has to be in between $\sim$0.1 and $\sim$1 and that it keeps roughly constant in the JED. Now, how is $\mu$ expected to vary in a SAD? Assuming that, in steady state, the poloidal field is mostly vertical i.e. no significant bending within the SAD, the radial distribution $B_z(r)$ in the SAD  is provided by the induction equation:  
\begin{equation}
\nu_m \frac{\partial B_z}{\partial r} \simeq u_r B_z
\end{equation}
where $\nu_m$ is the turbulent magnetic diffusivity. This equation has the obvious  {\it exact} solution 
\begin{equation}
B_z \propto  r^{- {\cal R}_m}
\label{eq:Bz}
\end{equation}
where ${\cal R}_m = -r u_r/\nu_m$ is the (effective) magnetic
Reynolds number. In a turbulent disc one usually assumes that all anomalous transport
coefficients are of the same order so that $\nu_m \simeq \nu_v$, $\nu_v$
being the turbulent viscosity. Since the (effective) Reynolds number
${\cal R}_e = -r u_r/\nu_v = 3/2$ in a SAD, one gets that any vertical
magnetic field is naturally {\it increasing} towards the center. Now in a
SAD of vertical scale height $h(r) \propto r^\delta$, the total pressure
$P_{tot} = \rho \Omega_k^2 h^2 $  scales as
\begin{equation}
P_{tot} = \frac{\dot M_{SAD} \Omega_k^2 h}{6\pi \nu_v} \propto r^{-3/2 - \delta} 
\end{equation}
where $\dot M_{SAD}$ is the (constant) SAD accretion rate. Using Eq.~(\ref{eq:Bz}) we get 
$\mu \propto r^{-\xi} \, \, \mbox{      with      }\, \,  \xi = 2 {\cal R}_m - \delta - 3/2$.
In a SAD ${\cal R}_m \simeq 3/2$ and $\delta$ is always close to unity (apart from the unstable radiation pressure dominated zone where $\delta=0$). Of course, the real value of $\xi$ critically depends on the magnetic Prandtl number (${\cal P}_m = \nu_v/\nu_m$) but this result suggests that, in a SAD, one may reasonably expect $\mu$ to increase towards the center. In conclusion, a reasonable radial dependence of $\mu$ should look like the one plotted on top of Fig. \ref{muprofil} i.e. constant (but in the range 0.1--1) between $r_{in}$ and $r_{tr}$ in the JED and decreasing above $r_{tr}$ in the SAD.\\

The limited range of values of $\mu$ in a JED bring us to the following assumptions. First, at a given radius in the disc, we assume that:
\begin{itemize}
\item  a SAD-to-JED transition occurs when $\mu$=1
\item a JED-to-SAD transition occurs when $\mu <$0.1
\end{itemize}
On the other hand, $\mu\propto\displaystyle\frac{B^2}{\mdot}$, meaning that $\mu$ is controlled by $\mdot$ but also by the magnetic field strength. This ratio is not expected to be constant with $\mdot$ and we make the zero order approximation that it anti-correlates with the accretion rate i.e. we assume that  the variation of the accretion rate is the dominant cause of the variation of $\mu$. As we will see below, these different assumptions provide a simple way to explain the hysteresis behavior observed in microquasars.\\

 We have reported in Fig. \ref{muprofil} the light curve and hardness-intensity diagram of GX 339-4 during its 2002/2003 outburst \cite{bel05} as well as the expected evolution of the radial distribution of the disc magnetization $\mu$. Starting from  the initial quiescent state in A, where $\mu$ is constant equal to 1 in the JED  and decreases in the SAD, we reach the highest flux level hard state B after a large increase of the accretion rate. Following our assumptions, from A to B,  $\mu(r)$ simply decreases but keeps its global shape constant. Point B corresponds to the situation where $\mu(r_{tr})=$0.1. A small increase of the accretion rate then produces a JED-to-SAD transition starting from the outer part of the JED, i.e. the transition radius $r_{tr}$ decreases. The system then reaches point C. This may correspond to two different situations. Either $r_{tr}$ reaches a critical values that switch on a pair cascade process giving birth to ultra-relativistic pair blobs, or $r_{tr}$ becomes to close to $r_{in}$ for the MHD jet structure to keep stable. Both cases lead to the same effect i.e. the disappearance of the MHD jet. In the former case, the pair blobs produce strong radio emission and cause possible superluminal events while in the latter case the transition should be less "explosive". In any case, the situation in C is expected to be quite unstable, the MHD jet structure oscillating between life or death while some pair blobs are randomly produced. As already said in Sect. \ref{evolution} this could explain the complex spectral and temporal behavior that characterize the upper left corner of the HID diagram. Anyway, from C to D we enter in a disc dominated state. Whatever the accretion rate does, and following our assumptions, the JED cannot re-appears unless $\mu(r_{in})$ becomes equal to 1. In the case of GX 339-4, this corresponds to point D where $\mdot$ becomes sufficiently small for this condition to be fulfilled. From point D to A, a SAD-to-JED transition occurs starting from $r_{in}$ on and in A we turn back to the initial conditions, at the beginning of the outburst. Thus the hysteresis behavior is a direct consequence of the variation of $\mu$ during the outburst, variation controlled by the accretion rate.

It is worth noting here that, due to the  unknown dependence of the magnetic field with the accretion rate, SAD-to-JED and JED-to-SAD transitions are not expected to occur at the same accretion rate for different objects. Even in a given object the transitions can be different for different outbursts  (like in the case of GX 339-4, cf. \cite{bel06}) since it depends certainly  on the past history of the accreting system.

\section{What about the X-ray corona?}
\label{sectcorona}
\begin{figure*}[b]
\centering
\begin{tabular}{cc}
\includegraphics[width=0.35\textwidth]{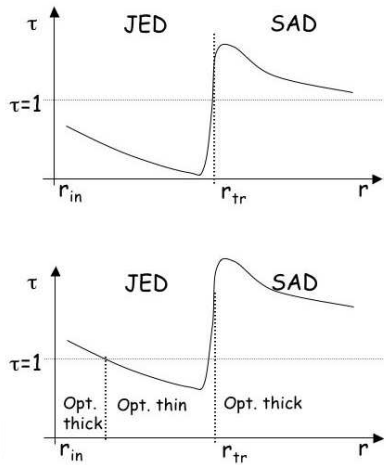}&
\begin{minipage}{0.45\textwidth}
\vspace*{-7cm}
\caption{{\bf Top:} The expected radial distribution of the vertical optical depth in the disc for a relatively low accretion rate. It strongly decreases inward at the transition between the SAD and the JED due to the strong decrease of the surface density in the JED (see Sect. 4).  The JED is optically thin. {\bf Bottom:} the same for a larger accretion rate. The middle part of the JED is optically thin while the inner and outer parts are optically thick. The optically thin region in the JED is expected to play the role of the X-ray "corona". For even higher accretion rate, the entire JED is expected to be optically thick. \label{distrtau}}
\end{minipage}\\
\end{tabular}
\end{figure*}
The radio/X-ray correlations found in XrB during their hard states (e.g. \cite{cor00,cor03,gal03} imply a strong link between  the hard X-ray corona  and the radio jet. It is now commonly believed that the corona is indeed the base of the jet and we show in this section that this corona is more likely a part of the JED itself. Indeed, as already said, self-consistent accretion-ejection simulations show that the magnetic field has to be close to equipartition with the total pressure in the JED i.e. $v_A\sim c_s$ where $v_A$ and $c_s$ are the Alfven and sound velocity respectively. In the jet however  $v_A$ is close to the speed of light (but still midly relativistic). The JED and the jet basis being threaded by the same magnetic field, this implies that the jet to disc density  $\displaystyle\frac{n_{jet}}{n_{JED}}<\left (\frac{h_{JED}}{r}\right ) ^2$
where $h_{JED}$ is the JED height scale. This is a relatively important constraint since simulations give at most $\varepsilon=\displaystyle\frac{h_{JED}}{r}\lta 0.1$ and is more likely $< 0.01$ . Thus the jump in density between the JED and the jet basis can be very large. Then it can be shown that the vertical optical depth of the jet basis is at most equal to:
\begin{equation}
\tau_{base}=3 \times 10^{-4} \left ( \frac{\dot m}{0.01}\right )\left ( \frac{r}{10}\right )^{-\frac{3}{2}}\left (\frac{H_{base}}{10 r_G} \right )
\end{equation} 
where  $\dot m = \dot M c^2/L_{Edd}$ and $H_{base}$ the height scale of the jet basis (in physical units). However fits of the X-ray spectra in the hard state require generally a corona optical depth of $\sim$ 1 which implies too large jet basis size, completely inconsistent with observations (for example concerning the X-ray variability).


Consequently, the hard X-ray "corona"  is more likely in the JED itself. This is not unreasonable since we expect optically thin regions in the JED even at high accretion rate. This is a direct consequence of the torque produces by the jet in the accretion disc. First most of the accretion power goes in the jet so the protons in the disc are "cold" and the JED keeps geometrically thin (or at most slim). Secondly, the accretion velocity in the JED is much larger than in a SAD. Simulations show that $\displaystyle u_r^{JED}\simeq \frac{r}{h_{JED}}u_r^{SAD}\gg u_r^{SAD}$ then implying  the vertical optical depth $\tau_{JED}\ll\tau_{SAD}$. In conclusion, the radial profile of the vertical optical depth in the disc should look like Fig. \ref{distrtau}. For low accretion rate, the entire JED can be optically thin. For higher $\mdot$, situations can be reached where the JED is optically thin in its middle part but optically thick in its inner (close to $r_{in}$) and outer (close to $r_{tr}$) parts. The last situation is very interesting since it could explain the presence of broad iron lines observed some times in  hard state (e.g. \cite{mil02,mil05}) and requiring the presence of optically thick material close to the black hole. Work is in progress to better quantify these results.
 
\section{A toy model for the X-ray corona in the Hard state}

\begin{figure*}[b]
\centering
\begin{tabular}{cc}
\includegraphics[width=0.45\textwidth]{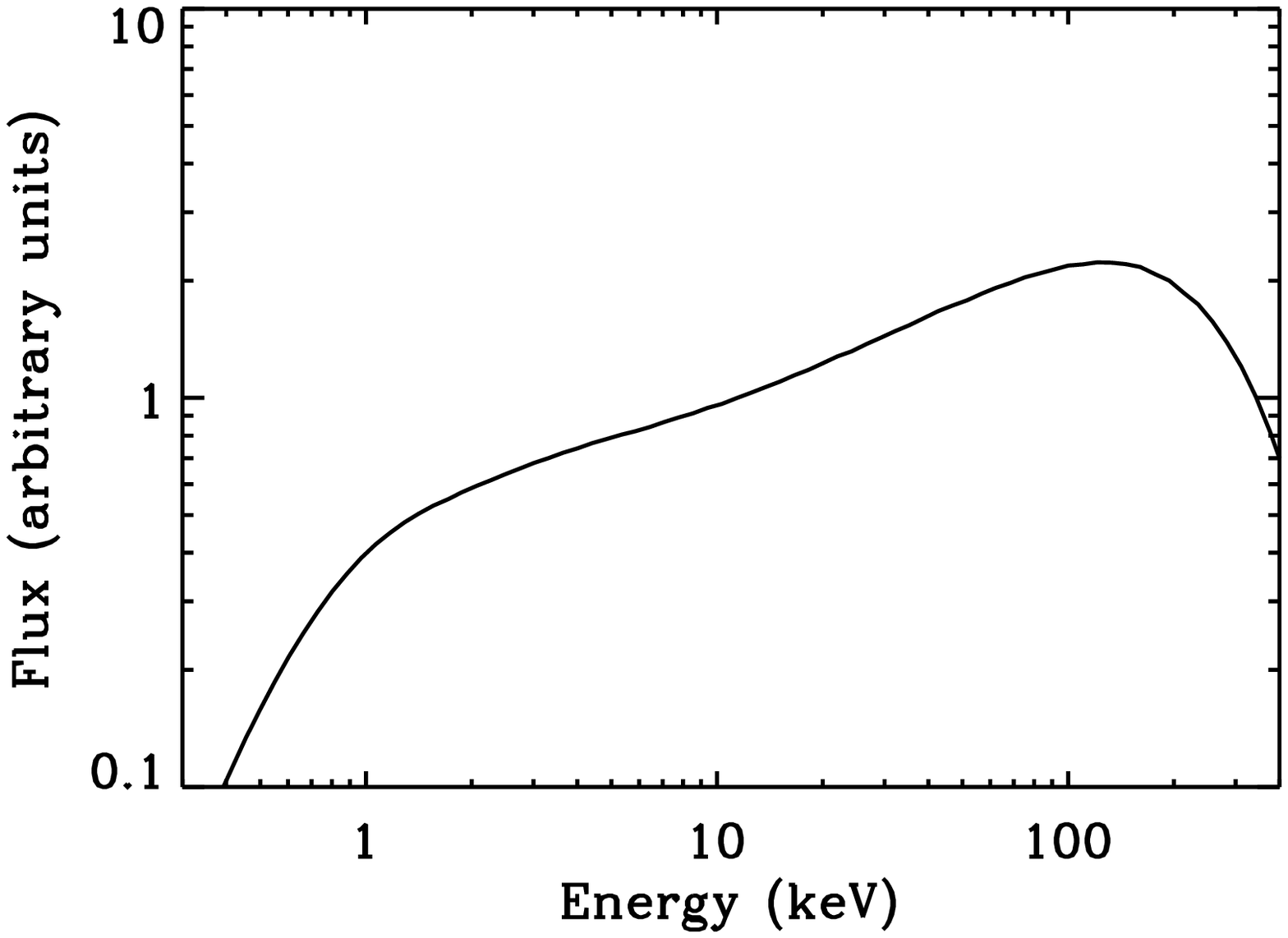}&
\includegraphics[width=0.45\textwidth]{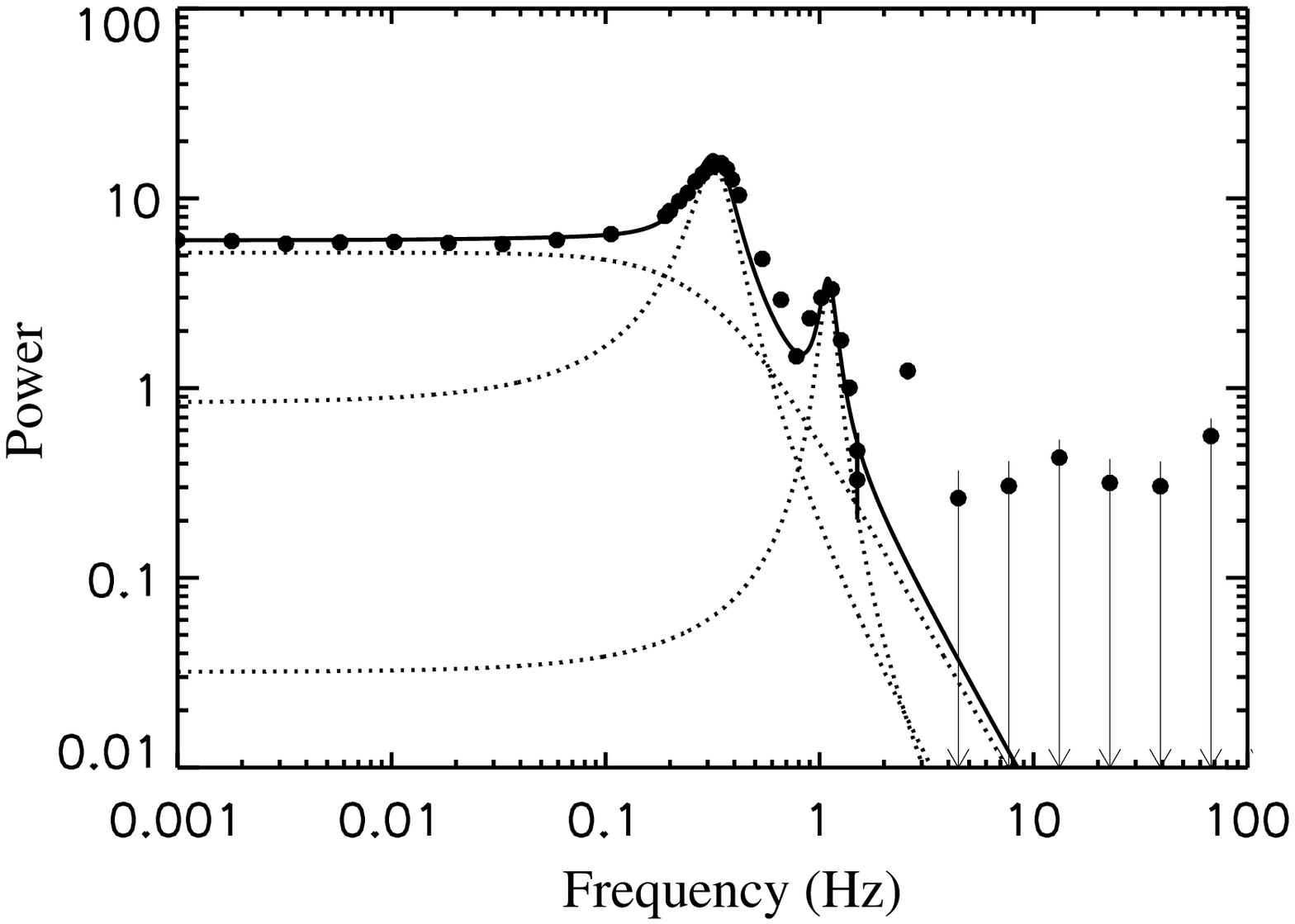}
\end{tabular}
\caption{Simulated Monte Carlo SED and PDS
expected from 
a cylindrical corona with mean optical depth and temperature $\tau=1.4$
and $kT_e= 100$ keV when this corona is excited by a external white noise. In the PDS, we have added 3 different lorentzians to fit the broad band noise and the two major QPOs. 
}
\label{fig:sedpds}
\end{figure*}
We have also investigated the origin of the high energy variability in microquasars focusing first on the Hard State variability. In our case (see previous section) this corona should look like a cylinder in the JED and we simply assume in this section that $r_{in}$
and $r_{tr}$ are its inner and outer radii respectively. The out-coming
spectrum then depends on the optical depth and temperature of this
corona. The observed spectra in the hard state generally required
$\tau\sim 1$ and $kT_e\sim 100$ keV. We propose that the observed
variability in this state is produce by the oscillation of the corona
through external excitation in $r_{tr}$ producing compressive waves in
the corona. This idea was already proposed by \cite{mis00} and we study more generally this problem through Monte-Carlo simulations. Here we focus on stationary compressive waves, the perturbations being totally reflected in $r_{in}$.  Perturbations $\partial n(r,t)$ and $\partial T_e(r,t)$ of
the corona density and temperature then produce spectral variability
in the X-ray spectra. For low frequency oscillations, the density and
temperature perturbation have the same sign in the corona. Thus we
expect large effects on the output spectrum i.e. large rms. On the
contrary, for high frequency oscillations,  $\partial n(r,t)$ and $\partial T_e(r,t)$  change sign  between $r_{in}$ and
$r_{tr}$. In this case, multiple Compton scatterings smear the spectral variability effects on the out-coming spectrum. The larger the oscillation frequency the weaker the variability and the smaller the rms. Now, depending on the boundary conditions, resonances where
$\partial n(r,t)$ and $\partial T_e(r,t)$ reach large amplitude may exist producing QPOs in the Power Density Spectrum (PDS). We show in Fig. \ref{fig:sedpds} the simulated Monte Carlo SED and PDS
expected from 
a cylindrical corona with mean optical depth and temperature $\tau=1.4$
and $kT_e= 100$ keV when this corona is excited by a external white noise and when $r_{in}$ and $r_{tr}$ are nodes of the compressive waves. The SED looks very similar to the common hard state SED. The PDS show
 a flat top profile at low frequency, decreasing at high frequency (above $\sim (kT_e/m_p)^{0.5}/ r_{tr}$), in agreement with the PDS observed in this state. However different QPOs corresponding to the different resonances of the system are also visible. While low frequency QPOs are known to be present in the hard state/High intermediate state, close to the cut-off frequency, the presence of multiple low frequency QPOs like in our simulations are rather uncommon (but see for example \cite{rod04b}). Works are in progress to include effects that could attenuate these QPOs  (like wave transmission in $r_{in}$, Cabanac et al. in preparation).




\end{document}